\documentclass[pre,twocolumn,amsmath,amssymb,superscriptaddress,showpacs]{revtex4-1}

\usepackage{amsmath,amssymb,graphicx}
\usepackage{amsbsy}
\usepackage{latexsym}
\usepackage{color}
\usepackage{graphicx}
\usepackage{psfrag}
\usepackage[normalem]{ulem}
\usepackage{bm}

\newcommand{\be}{\begin{equation}}
\newcommand{\ee}{\end{equation}}
\newcommand{\bea}{\begin{eqnarray}}
\newcommand{\eea}{\end{eqnarray}}
\newcommand{\bq}{\mathbf{q}}
\newcommand{\bx}{\mathbf{x}}
\newcommand{\br}{\mathbf{r}}

\newcommand{\comment}[1]{}

\begin{document}

\title{Quasi-deterministic dynamics, memory effects, and lack of self-averaging\\ 
in the relaxation of quenched ferromagnets}

\author{Federico Corberi}
\email{corberi@sa.infn.it}
\affiliation{Dipartimento di Fisica ``E.~R. Caianiello'', and INFN, Gruppo Collegato di Salerno,
and CNISM, Unit\`a di Salerno,Universit\`a  di Salerno, via Giovanni Paolo II 132, 84084 Fisciano (SA), Italy.}

\author{Eugenio Lippiello}
\email{eugenio.lippiello@unicampania.it}
\affiliation{Dipartimento di Matematica e Fisica, Universit\`a della Campania,
Viale Lincoln 5, 81100, Caserta, Italy}

\author{Paolo Politi}
\email{paolo.politi@cnr.it}
\affiliation{Istituto dei Sistemi Complessi, Consiglio Nazionale delle Ricerche, Via Madonna del Piano 10, 50019 Sesto Fiorentino, Italy}
\affiliation{INFN Sezione di Firenze, via G. Sansone 1, 50019 Sesto Fiorentino, Italy}

\begin{abstract}

We discuss the interplay between the degree of dynamical stochasticity, memory persistence and
violation of the self-averaging property in the aging kinetics of quenched ferromagnets.
We show that, in general, the longest possible memory effects, which correspond to the slowest possible temporal
decay of the correlation function, are accompanied by the largest possible violation
of self-averaging and a quasi-deterministic descent into the ergodic components. 
This phenomenon is observed in different systems, such as the Ising model with long-range
interactions, including mean-field, and the short-range random field Ising model.
   
\end{abstract}

\maketitle

{\it Introduction ---}
When computing thermodynamic properties one must, in principle,
consider the full statistical-mechanical average $\langle \cdot \rangle$, 
namely over the realisations of the stochastic trajectories, the initial conditions and,
if present, over the quenched disorder distribution. 
However, if the sample has specific self-averaging properties, 
the latter two averages are  not necessary because they are
realised by the system itself in the thermodynamic limit.
Restricting for the moment the discussion to clean samples, 
i.e. without quenched disorder, this occurs when 
the system is ergodic. In this case after some time a large part of
phase space is visited, and the memory of the initial condition is fully lost:
therefore, the fate of a thermodynamical process does not depend on the
specific initial microstate belonging to the same macrostate. 

The situation is more subtle when phase space breaks into 
ergodic components~\cite{Palmer1982}, namely mutually non-accessible
regions. 
In this case, if the initial state is well inside one of such
components its memory cannot be deleted because the other cannot be accessed.
This is trivial for a uniaxial ferromagnet 
below the critical temperature $T_c$, where
the equilibrium magnetisation $M$ takes the two possible values $M_\pm =\pm M_{eq}$. 
A sample prepared with a macroscopic 
$M(t=0) >0$ ($<0$) evolves towards the positive (negative) equilibrium value
and self-averaging is not operating. 

A different situation occurs when the system is initially
on the boundary ${\cal B}$ between ergodic components. 
In ferromagnets, ${\cal B}$ is the set of configurations with $M\simeq 0$, and 
this happens when the initial state is sampled
from a high temperature ($T\ge T_c$) equilibrium state.
The evolution in this case proceeds by 
coarsening of domains of the competing equilibrium phases~\cite{CorPol15},
whose typical size $L(t)$, at time $t$, grows unbounded.
Aging is manifested~\footnote{However,
aging can be observed also in the absence of ergodicity breaking as, for instance,
in the case of a quench to a critical temperature.}  and the dynamics remains
on ${\cal B}$ for ever.
This is strictly true if the thermodynamic limit is taken 
before letting time $t$ large. However, in all physical situations, one deals with a large
but finite system. Therefore the initial state, due to thermal fluctuations, will have some
offset $M(0)$ from ${\cal B}$ and one can ask how this may change the destiny of the system.

The different options can be appreciated 
\comment{introducing the autocorrelation exponent $\lambda $ through $M(t)\sim L(t)^{d-\lambda}$,  where $d$ is the spatial dimension and $\lambda$
is bounded by the Fisher-Huse inequality~\cite{Fisher88,Yeung96}}
in terms of the exponent $\lambda$ controlling the decay of the autocorrelation function and also related~\cite{BrayDerrida1995} to the growth in time of the magnetization $M(t)\sim L(t)^{d-\lambda}$,
where $d$ is the spatial dimension. 
The Fisher-Huse inequality~\cite{Fisher88,Yeung96} fixes the bounds for $\lambda$
\be
\frac{d}{2}\le \lambda \le d .
\label{FHbounds}
\ee
If the system stays close to ${\cal B}$ for ever
~\footnote{In principle the duration of the process cannot extend forever due to
the finiteness of the system. Here however we are not considering this kind of finite-size effect} 
the magnetization does not amplify ($\lambda =d$), self-averaging is at work, 
and memory of the initial condition is retained the least possible~\footnote{In this paper 
the term {\it memory effects} refer to a persisting correlation of the system with the initial state, at variance with
the acceptation in glassy literature~\cite{Bouchaud2000} where it refers to the non equilibrium history of the system.}. 
In the opposite situation 
the system deterministically falls in the ergodic component selected by the sign of $M(0)$. 
In this case the offset $M(0)$ is strongly amplified and $M(t)$ grows as fast as possible, i.e. $\lambda=d/2$. 
This process is associated to the longest possible memory of the initial condition and to the strongest
violation of self-averaging. In between these two extrema 
there is a continuum of options, with $d/2<\lambda<d$.

Existing analytical~\cite{Kissner_1993,PhysRevB.44.9185,Mazenko98,PhysRevE.65.046136} 
and 
numerical~\cite{Lorenz_2007,PhysRevE.68.065101,Abriet2004,PhysRevB.42.4514,Bray_1990,PhysRevB.44.9185} 
determinations of $\lambda$
suggest that the maximum of memory, $\lambda=d/2$, 
is only approached in unphysical limits, diverging space 
dimension limit $d\to \infty$ or diverging order parameter components limit ${\cal N}\to \infty$.
Instead, upon associating the origin of the lower bound $\lambda=d/2$ to 
some deterministic properties of the dynamics, in this paper we show that 
it is possible to toggle among all the three situations above and that the case 
with $\lambda=d/2$ is found also for finite $d$ and ${\cal N}$
in the presence of long-range interactions or in the presence of quenched disorder. 
\comment{A similar effect can be obtained in systems
with short-range interactions by adding 
quenched disorder.}   
 
{\it The model and the two limiting regimes ---}
In order to set the stage with a specific example, let us start our discussion 
by considering the one-dimensional clean ferromagnet described by the Hamiltonian
\be
H= -\frac{1}{2}\sum_{i,j} J(|i-j|)  s_i s_j ,
\label{e.H}
\ee 
where $s_i=\pm 1$ are $N$ Ising variables, and $J(r)= \delta_{r,1}$ for nearest
neighborgs (nn) couplings, and $J(r)=1/r^{1+\sigma}$ in the case of 
long-range interactions. We will focus on the case $\sigma >0$ where 
additivity and extensivity hold~\cite{review_long_range}.
The model has a ferromagnetic phase below a finite critical temperature 
$T_c(\sigma) >0$ for $\sigma < 1$~\cite{Dyson1969,Tomita};
it has a Kosterlitz-Thouless transition~\cite{Frohlich1982} for $\sigma=1$;
finally, $T_c=0$ for $\sigma > 1$.

Let us now discuss the relaxation of the model with a non conserved order parameter
after a quench from $T_i=\infty$ to a low $T$.
We consider Glauber dynamics where a random
spin is reversed with probability 
$w=(1+\exp(\Delta E/T))^{-1}$, where 
$\Delta E$ is the energy difference due to the spin-flip.
Not only the static properties, also the non-equilibrium kinetics
change crossing $\sigma =1$.
$L(t)\sim t^{1/z}$ grows with 
a dynamical exponent~\cite{BrayRut94,RutBray94} $z=1+\sigma$ for
$0<\sigma\le 1$ or $z=2$ for $\sigma >1$ and nn.
This behavior is captured by a single domain model.
The distance $X(t)$ between two neighbouring domain walls satisfies an overdamped
Langevin equation, $\dot X(t) = - F(X) + \xi(t)$, where $F(X)$ is a force determined
by Eq.~(\ref{e.H}) and $\xi(t)$ is a gaussian white noise.
The force is given by $F(X)=-U'(X)$, where $U(X) = \sum_{i=1}^X \left(\sum_{j=-\infty}^0
+\sum_{j=X+1}^\infty\right) J(|i-j|)$. 
For large $X$ we can replace discrete summations with integrals and
evaluating the integrals in the brackets we obtain $U(X) \simeq (2/\sigma) \int_1^X ds/s^\sigma$, therefore
$F(X) \sim -1/X^{\sigma}$. Given that $F(X)$ is the average speed of the domain wall the
closure time of a domain of initial size 
$X(0)=L$ is $t=\int _L^0 dX/F(X) \propto L^z$ with $z=1+\sigma$ for $\sigma \le 1$ and $z=2$ for $\sigma>1$.
The difference between these two regimes is due to
the deterministic force $F(X)$, that affects the coarsening process in the former ($\sigma\le 1$)
while it is irrelevant in the latter ($\sigma>1$). For this reason these regimes will be called {\it convective}
and {\it diffusive} regimes, respectively.

\begin{figure}
\begin{center}
\includegraphics[width=0.45\textwidth,clip]{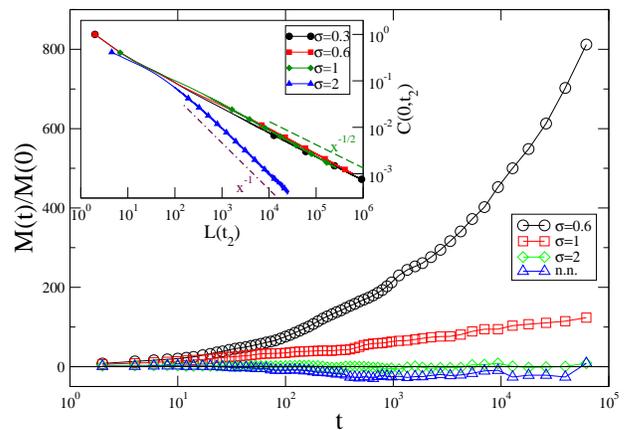}
\end{center}
\caption{The fluctuating magnetization $M(t)$ for a single realization starting from the same initial 
condition. The system size is $N=10^6$ and the quench temperature is $T=0.1$.
In the inset $C(0,t_2)$ is plotted against $L(t_2)$  
for different $\sigma$ after a quench to $T=0.1$. 
System size is $N=2\times 10^7$.
The dashed straight lines are the
decays $x^{-\lambda}$ with $\lambda=1$ and $\lambda=1/2$.}
\label{f.examples}
\end{figure}

These two regimes can be clearly distinguished
by considering the fluctuating magnetisation $M(t)=\sum_{i=1}^N s_i$,
which is shown in Fig.~\ref{f.examples} 
for systems prepared 
with a fixed condition $M(0)$ $\sim \sqrt{N}$ equal for all $\sigma$-values.
In the convective regime $M(t)$ asymptotically diverges and it typically has the same
sign as $M(0)$~\footnote{In our simulations this occurs in the $\sim 80\%$ of the thermal histories.
This fraction does not seem to depend significantly on the size $N$ of the system.
What is mostly important is that  for {\it all} thermal histories
there exists a finite time $t^*$ beyond which $M(t)$ does not change sign and whose
absolute value increases in time}.
In the diffusive regime it fluctuates around $M(0)$.
This means that the convective regime keeps memory of the initial condition, while 
the diffusive does not. This implies that decorrelation is slower in the first case and,
actually, we will show in a moment that it occurs in the slowest possible way. 
Self-averaging with respect
to initial conditions is broken for $0<\sigma \le 1$ (convective regime) 
while it holds for $\sigma >1$ (diffusive regime).

With this example in mind, we now turn to a more general discussion.
Let us consider the correlation function which, 
using a continuous picture for a scalar field~\footnote{Our results can be straightforwardly 
generalized to a vector order parameter}  $\phi(\bx,t)$, reads
$S(r;t_1,t_2) \equiv \langle \phi(\bx+\br,t_1) \phi(\bx,t_2)\rangle $, 
where $t_2 > t_1$ and
$\langle\cdots\rangle$ is the full non-equilibrium statistical average.
We focus on the scaling regime where the autocorrelation function $C(t_1,t_2)=S(r=0;t_1,t_2)$ 
behaves as~\cite{Bray94}
\be
C(t_1,t_2)
\simeq \left [L(t_1)/L(t_2)\right ]^\lambda,
\label{e.scalingC}
\ee
where, as it will be discussed around Eq.~(\ref{e.M1M2}), $\lambda $ is the same exponent introduced before
which therefore obeys Eq.~(\ref{FHbounds}). 

{\it The inequalities for $\lambda$ ---}
A derivation of Eq.~(\ref{FHbounds}) is now provided following~\cite{Yeung96}.
We indicate with $u_l = \phi_l(\bq,t_1)$ the Fourier transform of the field  
$\phi_l(\bx,t_1)$ evaluated at the time $t_1$ during the $l$-th realization of the dynamics.
Similarly we define $v_l = \phi_l(\bq,t_2)$ at the time $t_2$.
We can therefore define the scalar product as
$\vec u\cdot \vec v \equiv  (2\tilde N)^{-1}\sum_l( u_l v_l^* + c.c)
= \frac{1}{2} \left[ S(q,t_1,t_2) + S^*(q,t_1,t_2) \right]$,
where $\tilde N$ is the number of realizations and 
$S(q,t_1,t_2) \equiv \langle \phi(\bq,t_1) \phi(-\bq,t_2) \rangle$
is the Fourier transform of $VS(r;t_1,t_2)$, $V$ being the system volume. 
We can now apply the Cauchy-Schwarz inequality,
$| \vec u\cdot \vec v | \le |u||v|$
and obtain
\be
\frac{1}{2} | S(q,t_1,t_2) + S^*(q,t_1,t_2) | \le \sqrt{S(q,t_1) S(q,t_2)},
\label{eq.S}
\ee
where, for ease of notation, $S(q,t) \equiv S(q,t,t)$.
If we integrate over $\bq$ we find
\be
C(t_1,t_2) \le \frac{1}{V(2\pi)^d}  \int d\bq \,\sqrt{S(q,t_1) S(q,t_2)} .
\label{eq.C}
\ee
Using Eq.~(\ref{e.scalingC}) and the scaling form $S(q,t) = L^d(t) f(qL)$,
with $f(x) \simeq 1$ for $x\ll 1$ and $f(x)$ negligibly small for $x\gg 1$,
we find the lower bound of Eq.~(\ref{FHbounds})~\footnote{We do not
consider the case of critical quenching}.

We now originally prove that the same lower bound can be derived from the term $q=0$ only of Eq.~(\ref{eq.S}),
\be
S(0,t_1,t_2) \le \sqrt{S(0,t_1) S(0,t_2)} \, .
\label{e.S0}
\ee
Using the scaling form for $S(q,t)$ (see below Eq.~(\ref{eq.C}))
it is straightforward to rewrite the previous equation as
\be
S(0,t_1,t_2) \le f(0)(L_1 L_2)^{d/2},
\label{e.S02}
\ee
where we used the shorthand $L_1 \equiv L(t_1)$, and similarly for $L_2$.
The left-hand side of Eq.~(\ref{eq.C}) can be worked out expressing the two-time correlation function as follows,
$C(t_1,t_2) = \frac{1}{V(2\pi)^d}\int d\bq \,S(q,t_1,t_2) 
= \frac{L_2^d}{V(2\pi)^d}\int d\bq \, F(qL_2,L_1/L_2) ,\nonumber$
where we have used the scaling hypothesis
$S(q,t_1,t_2)=L_2^d F(qL_2,L_1/L_2)$,
valid when both times $t_1$ and $t_2$ are in the scaling regime.
In the limit of large $L_2$ (i.e., of large $t_2$) only wavevectors
$q<1/L_2$ contribute to the integral. If $S(q\to 0,t_1,t_2)$ goes to a constant,
which is the case for quenches below $T_c$ or to $T=0$, we can finally write
\be
C(t_1,t_2) \simeq \frac{1}{V(2\pi)^d} \frac{S(0,t_1,t_2)}{L_2^d}.
\label{eqeq}
\ee
Using this relation and Eq.~(\ref{e.S02}) we find
$C(t_1,t_2) \le \mbox{const}\times (L_1/L_2)^{d/2}$
and the scaling form  (\ref{e.scalingC}) gives $\lambda \ge d/2$.
Therefore Eq.~(\ref{e.S0}) is equivalent to the lower bound~(\ref{FHbounds}).

\comment{We remark that the lower limit in the Cauchy-Schwarz inequality is achieved when 
the vectors $u_l$ and $v_l$ are parallel. This can occur when, during the same realization of the dynamics, the configuration at time $t_1$ determines the subsequent one as also highlighted by the behavior of the magnetization.} 

The upper bound in Eq.~(\ref{FHbounds})
is defined in~\cite{Fisher88} as a ``suggestive bound" because it cannot be
proved as rigorously as the lower bound.
In order to derive it,  starting from the straightforward relation
$\langle M(t_1) M(t_2)\rangle = S(0,t_1,t_2)$,
and using Eqs.~(\ref{eqeq}) and (\ref{e.scalingC}), one arrives at
\be
\langle M(t_1) M(t_2)\rangle = \mbox{const}\times L_1^\lambda L_2^{d-\lambda} .
\label{e.M1M2}
\ee
Authors of Ref.~\cite{Fisher88} argue that $\lambda \le d$ because ``forgetting of an initial bias 
appears unlikely".
In other words the strongest memory loss corresponds to the limit $\lambda = d$.

{\it Averaging and memory ---}
We now consider the role of the different statistical averages.
The full one $\langle \cdots \rangle$ 
is taken over the stochastic trajectories, $\langle \cdots \rangle_{tr}$; 
the initial condition, $\langle \cdots \rangle_i$; and, if present,
over the quenched disorder, $\langle \cdots \rangle _q$.
Let us consider, to begin with, a clean system.
We can split the fluctuating magnetisation as 
$M(t)=\langle M(t)\rangle_{tr}+\psi(t)$, where $\psi(t)$ is the stochasticity 
left over after taking the partial averaging $\langle M(t)\rangle_{tr}$, 
so that $\langle \psi(t)\rangle_{tr}\equiv 0$.
Then we have
$\langle M(t_1)M(t_2)\rangle =\big \langle \langle M(t_1)\rangle _{tr}\langle M(t_2)\rangle_{tr}\big \rangle_i
+\big \langle \psi(t_1)\psi(t_2)\big \rangle$.
If we now fix $t_1$ and let $t_2$ diverge, 
$\langle \psi(t_1)\psi(t_2) \rangle = \langle \psi(t_1)\rangle \langle \psi(t_2)\rangle =0$
and from Eq.~(\ref{e.M1M2}) we obtain
\be
\big \langle \langle M(t_1)\rangle_{tr}\langle M(t_2)\rangle _{tr}\,\big \rangle_i \simeq  L_2^{d - \lambda} .
\label{eqmm}
\ee
Next we argue that, if the quench is made in a ferromagnetic phase,
due to the presence of two ergodic components, for large
$t_1$ it is $\mbox{sign} \big( M(t_1)\big) =\mbox{sign} \big( M(t_2)\big)$. This is very well observed
for $\sigma <1$, see Fig.~\ref{f.examples}. 
Hence it is also $\mbox{sign} \langle M(t_1)\rangle_{tr}=\mbox{sign} \langle M(t_2)\rangle_{tr}$,
therefore Eq.~(\ref{eqmm}) (valid for $t_1$ fixed) amounts to
\be
\langle M(t)\rangle _{tr} \simeq  L(t)^{d - \lambda},
\label{e.Mt}
\ee
where we have denoted $t_2$ as $t$ to ease the notation.
Notice that the equation above is more general and applies to systems without a proper
ferromagnetic phase, such as  the $1d$ Ising model with
$\sigma >1$ or with nn,
because in this case there is no development of magnetisation starting
from a given state, see Fig.~\ref{f.examples}, and indeed
it is $\lambda =d$. 

Equation~(\ref{e.Mt}) shows that the slowest possible decorrelation, 
$\lambda=d/2$, is accompanied by the
fastest possible growth of the magnetization developed from an initial 
condition~\cite{BrayDerrida1995}. Let us observe that such maximum growth
is the one expected upon assuming a random
arrangements of a number $\sim VL^{-d}$ of domains of size $L$ 
each contributing a magnetisation $\sim L^d$. 
Eq.~(\ref{e.Mt}) for $\lambda=d/2$ then
derives from the central limit theorem.

The result~(\ref{e.Mt}) implies also that there is breaking of self-averaging with respect
to initial conditions if $\lambda <d$, as reflected by the fact that, for large 
$N$, the observable
{\it magnetisation} does not attain its average value $\lim _{N\to \infty} \langle M(t)\rangle =0$
unless the average over initial conditions is performed.
The most severe self-averaging breakdown occurs when $\lambda$ is 
at the lower bound in~(\ref{FHbounds}), whilst it is fully restored when
it is at the upper bound.
 
Let us put these arguments to the test in different models, starting
from the $1d$ model of Eq.~(\ref{e.H}). 
Let us recap what is known about $\lambda$.
For nn there is the
exact result~\cite{Glauber1963} $\lambda =1$, the upper bound of
Eq.~(\ref{FHbounds}) is saturated and self-averaging holds. 
For the long-range case it was shown in~\cite{CLP_lambda} that
there are two universality classes associated to the values $\lambda =1$ (for $\sigma >1$)
and $\lambda =1/2$ (for $\sigma \le1$). Since it is known~\cite{Yeung96}
that for non conserved order parameter this exponent is independent of $t_1$,
the best determination can be obtained by letting $t_1=0$. This is displayed
in the inset of Fig.~\ref{f.examples}, where $C(0,t_2)$ is shown for various choices
of $\sigma$, showing that $\lambda=1$ for $\sigma>1$ and $\lambda=1/2$ for $\sigma \le 1$. 

\begin{figure}
\begin{center}
\includegraphics[width=0.45\textwidth,clip]{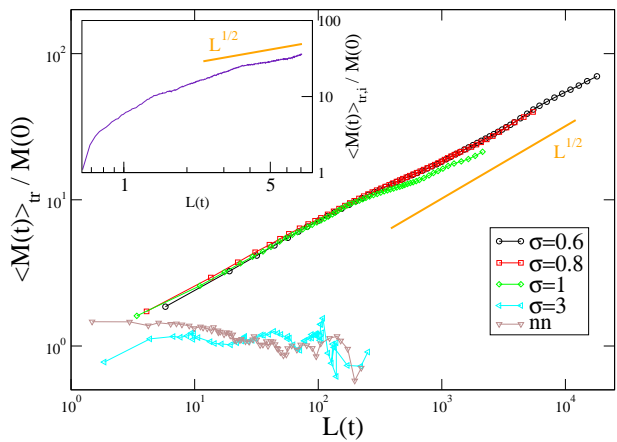}
\end{center}
\caption{$\langle M(t)\rangle_{tr}$ 
(normalized by its typical 
initial value $M(0)$) is plotted against $L(t)$ for 
	different $\sigma$. The system size is $N=10^6$ and the quenching temperature 
	is $T=0.1$. The orange straight line is the behavior $L(t)^{1/2}$.
	In the inset a similar plot is shown for $\langle M(t)\rangle_{tr,i}$ in the 
	$1d$ RFIM quenched to $T\to 0$ with $h/T=1/2$.  
          The system size is $N=10^5$.}
\label{f.dynamics}
\end{figure}

In Fig.~\ref{f.dynamics}  we plot
$\langle M(t)\rangle _{tr}$ as a function of $L(t)$, for different
$\sigma$ and {\it the same initial condition}. This shows very clearly that
in the convective regime ($0<\sigma\le 1$) where $\lambda =d/2$ it is
$\langle M(t)\rangle _{tr}\sim \sqrt{L(t)}$ while in the diffusive case ($\sigma >1$ or nn)
it is $\langle M(t)\rangle_{tr}\sim M(0)$, as expected after Eq.~(\ref{e.Mt}). 
Hence $\sigma =1$ separates the two opposite situations
in which the dynamics occurs on the boundary ${\cal B}$ of the ergodic components
(for $\sigma >1$) from the one where it deterministically sinks into such components (for $\sigma \le 1$). 
We should stress that $T_c=0$ is not a sufficient condition to have $ \lambda=d$,  
as attested by the $2d$ XY-model where 
$\lambda\simeq 1.17 < d=2$ even if $T_c=0$~\cite{1.17}.

In our model (\ref{e.H})
determinism can be ascribed to the convective character of domain wall motion~\cite{CLP_review,CLP_lambda}. 
Let us suppose to have two close domains of
sizes $\ell_1,\ell_2$, with $\ell_2$ slightly larger than $\ell_1$. 
In the diffusive case the average closure time of $\ell_1$, $\overline{t_1}$, is slightly smaller 
than the one of $\ell_2$, $\overline{t_2}$,
but the probability that $t_1 < t_2$ is only slightly larger than $1/2$.
In the convective regime, instead, the dominance of the deterministic force 
makes a domain wall always move towards the closest one~\cite{CLP_lambda},
so that $t_1$ is {\it always} smaller than $t_2$.
This induces a memory effect, since domains which are
eliminated have a larger probability to be anti-aligned with $M(t)$ and their removal 
further increase $M(t)$.  
Summarising, in the convective regime there is a reduced degree of stochasticity
and an increased memory 
with respect to the diffusive one, and this is the physical origin of the saturation of $\lambda $ 
to the lower bound.

{\it Same ideas, other models ---}
Let us now apply these ideas to different systems, starting from
the short-range ferromagnetic model in $d>1$.
In this case we have strict inequalities for any $d$, $d/2<\lambda <d$~\cite{Mazenko98}.
Hence self-averaging is spoiled, $\lambda<d$, in opposition to $d=1$.
This is because in $d>1$
interfaces do not freely diffuse, there is a deterministic drift induced by the curvature. However the
fate of the system is not fully determined by such deterministic force
because the shape of the percolating cluster plays a major role in the subsequent dynamics~\cite{BlaCorCugPic14}.
Hence there is only a weak drift from ${\cal B}$ towards the ergodic components and $\lambda $
stays larger than $d/2$.

When long-range interactions are present, results in $d>1$ are rare~\cite{Janke_lr_2019}
and studies of $\lambda$ are almost absent~\cite{Henkel_lr_preprint}.
However it is interesting that for the nn case in the limit $d\to \infty$,
which corresponds to the, so to say, {\it longest possible range of interactions}, the mean-field, 
one has $\lambda \to d/2$~\cite{Mazenko98} and $M(t)\sim L(t)^{d/2}$~\cite{CLP_epl,CLP_review}, as 
expected on the basis of our previous argument. 
In this limit there are no interfaces
and, therefore the strong memory effects leading to $\lambda =d/2$ cannot be associated
to the determinism of their motion, as in finite dimension. 
Instead, it can be observed that mean-field 
amounts to an averaging procedure which makes the
evolution, in a sense, more deterministic. Again, this reduction of the stochastic 
degree is perhaps the physical origin
of the saturation of $\lambda $ to the lower bound of Eq.~(\ref{FHbounds}). 

There is another well known limit in which phase-ordering has a similar character.  
This is the case of a vectorial order parameter 
$\vec \phi (\bx,t)$ with a large number ${\cal N}$ of components
and short-range interactions.
In the ${\cal N}\to \infty$ limit (a model sometimes denoted also as {\it spherical model}) one 
finds~\cite{PhysRevE.65.046136} $\lambda =d/2$ for any $d$~\footnote{The same is true for long range interactions provided that $\sigma <d$~\cite{CANNAS2001362}}.
By choosing an initially magnetised state it can be shown~\cite{JPhysA.26.3037}
that the magnetisation evolves deterministically as
$M(t)\sim L(t)^{d/2}$, as expected after Eq.~(\ref{e.Mt}).
It must be recalled that the large-${\cal N}$ limit
effectively amounts to replace $\phi^2$ with its mean value~\cite{PhysRevE.65.046136}.
Then, similarly to mean-field, the model realises a sort of averaging
wich tames the stochasticity and sets $\lambda$ to the minimum possible value.

Up to now we have only considered clean systems. It is now interesting to discuss 
the case with quenched disorder focusing, as a
paradigm, on the Random Field Ising Model (RFIM). 
The RFIM Hamiltonian is given by
Eq.~(\ref{e.H}), plus a contribution $- \sum_i h_i s_i$ due to a quenched random external
field that in the following we will consider with zero average and bimodal distribution $h_i=\pm h$.
We will focus on the nn case.
In order to discuss the role of the different averages, as done before, we must
now take into account that in this case also the quenched one $\langle \cdots \rangle_q$
comes into the game. Splitting the magnetisation as $M(t)=\langle M(t) \rangle _{tr,i}+\psi(t)$,
similarly to what done previously for the clean case but where now $\langle \cdots \rangle_{tr,i}$
is a partial average taken over both dynamical trajectories and initial conditions, one can follow
the same line of reasoning as before, arriving at the same results, replacing everywhere 
$\langle M(t)\rangle_{tr}$ with $\langle M(t)\rangle_{tr,i}$.

Let us start discussing the case with $d=1$, 
for which some analytical arguments are available. The model is
characterised~\cite{Corberi_rf} by a value of $\lambda $ at its
minimum, $\lambda=1/2$. 
 Hence, one should expect $\langle M(t)\rangle_{tr,i}\sim L(t)^{1/2}$.
In the inset of Fig.~\ref{f.dynamics} we plot $\langle M(t)\rangle_{tr,i}$ versus the average size of domains
$L(t)$ (which grows as $(\ln t)^2$). 
The result nicely confirms our expectation. In this case the growth of 
$\langle M(t)\rangle_{tr,i}$ can be traced back to the fact that the sum of the random fields
in a given quenched realisation is of order $N^{-1/2}$ and, hence, there is an explicit breaking
of the up-down spin symmetry. 
Hence, here it is the random field
which causes the deterministic fall into the ergodic components. 
Interestingly, this effect seems not to be limited to one dimension.
For $d>1$ the RFIM can only be studied numerically. For $d=2$ one 
observes~\cite{CLMPZ12} that $\lambda =d/2=1$ is still at the lowest possible
value, as for $d=1$. This suggests that the mechanism found in $d=1$
might be a general feature with random fields. 

{\it Conclusions ---}
We have interpreted the exponent $\lambda$ and its bounds, $d/2 \le\lambda\le d$,
in terms of stochasticity, memory effects,
ergodicity breaking and self-averaging.
When $\lambda=d$ memory is lost as fast as possible, 
magnetisation does not develop, and there is no breaking of self-averaging.
This occurs, for instance
in the $1d$ Ising model with nn, or in the $2d$, ${\cal O}(2)$ model E~\cite{NamKimLee}.
When $\lambda=d/2$ memory is maintained as much as possible, 
magnetization grows as $M(t)\sim (L(t))^{d/2}$ and there is a strong breaking of self-averaging.
This occurs in the $1d$ long-range Ising model with $\sigma \le 1$, in the mean-field
and spherical model limits, in the RFIM.
Between the two limiting cases, a continuum exists.

\comment{
Memory, encoded in the two-time correlation function in terms of the 
$\lambda $ exponent, is lost as fast as possible -- compatibly with basic principles -- 
when the upper bound of the Fisher-Huse inequality is met. 
In this case magnetisation does not develop regardless of the intial
preparation of the system, $\langle M(t)\rangle _{tr} \simeq M(0)$, 
and there is no breaking of self-averaging namely, in a clean system, 
$\langle M(t)\rangle _{tr}=\langle M(t)\rangle$. Averaging over initial conditions
is, in these cases, pointless.
This occurs, for instance
in the $1d$ Ising model with nn, or in the $2d$, ${\cal O}(2)$ model E~\cite{NamKimLee}.
In all the other cases, namely when $\lambda <d$, there is a violation
of self-averaging which gets more severe as  $\lambda$ approaches the lower
bound $\lambda =d/2$ and memory extends in time. 
This, of course, does not necessarily implies that other observables 
may not self-average, but this cannot be taken for granted.  
The arguments presented in this paper are rather general for systems quenched
to a phase with ergodicity breaking. Therefore we expect them to apply also  
to long-range systems in $d>1$. 
To our knowledge, the only study in this case is a 
preprint~\cite{Henkel_lr_preprint} where, however, the authors claim that in $d=2$ the 
Fisher-Huse inequality is violated, a fact worth of further investigations. 
}

It would be interesting to check if some model contrasts these ideas, starting
from long-range systems in $d>1$~\cite{Henkel_lr_preprint}.
The case of aging without ergodicity breaking, as in the case of a ferromagnet
quenched to the critical temperature, is also another test bench where
the relation between stochasticity, memory effects and self-averaging 
ought to be considered.
In this case the Fisher-Huse lower bound generalises~\cite{Yeung96} to
$\lambda \ge (d+\beta)/2$, where $\beta$ is an exponent characterizing the small $q$
behavior of the structure factor.
It would be interesting to check if in this case it is still possible to relate the 
bounds on $\lambda$ to specific features of the dynamics.

We thank Jorge Kurchan for discussions.
E. Lippiello and P. Politi  acknowledge
support from the MIUR PRIN 2017 project 201798CZLJ.

\bibliography{long_range}

\end{document}